\documentclass[12pt]{article}
\usepackage{amssymb}
\usepackage{graphicx}
\renewcommand{\theequation}{\arabic{section}.\arabic{equation}}

\begin{document}



\def\a{\alpha}
\def\b{\beta}
\def\d{\delta}
\def\e{\epsilon}
\def\g{\gamma}
\def\h{\mathfrak{h}}
\def\k{\kappa}
\def\l{\lambda}
\def\o{\omega}
\def\p{\wp}
\def\r{\rho}
\def\t{\tau}
\def\s{\sigma}
\def\z{\zeta}
\def\x{\xi}
\def\V={{{\bf\rm{V}}}}
 \def\A{{\cal{A}}}
 \def\B{{\cal{B}}}
 \def\C{{\cal{C}}}
 \def\D{{\cal{D}}}
\def\G{\Gamma}
\def\K{{\cal{K}}}
\def\O{\Omega}
\def\R{\bar{R}}
\def\T{{\cal{T}}}
\def\L{\Lambda}
\def\U{U_q(sl_2)}
\def\E{E_{\tau,\eta}(sl_n)}
\def\Zb{\mathbb{Z}}
\def\Cb{\mathbb{C}}

\def\R{\overline{R}}

\newcommand{\be}{\begin{eqnarray}}
\newcommand{\ee}{\end{eqnarray}}
\def\beq{\begin{equation}}
\def\eeq{\end{equation}}
\def\bea{\begin{eqnarray}}
\def\eea{\end{eqnarray}}
\def\ba{\begin{array}}
\def\ea{\end{array}}
\def\no{\nonumber}
\def\le{\langle}
\def\re{\}}
\def\lt{\left}
\def\rt{\right}
\def\non{\nonumber}
\newcommand{\tr}{\mathop{\rm tr}\nolimits}
\newcommand{\sn}{\mathop{\rm sn}\nolimits}
\newcommand{\cn}{\mathop{\rm cn}\nolimits}
\newcommand{\dn}{\mathop{\rm dn}\nolimits}
\newcommand{\kk}{\kappa}
\newcommand{\id}{\mathbb{I}}
\newcommand{\sgn}{\mathop{\rm sgn}\nolimits}
\newcommand{\ch}{\mathop{\rm ch}\nolimits}
\newcommand{\sh}{\mathop{\rm sh}\nolimits}
\newcommand{\tnh}{\mathop{\rm tanh}\nolimits}
\newcommand{\cth}{\mathop{\rm coth}\nolimits}

\newtheorem{Theorem}{Theorem}
\newtheorem{Definition}{Definition}
\newtheorem{Proposition}{Proposition}
\newtheorem{Lemma}{Lemma}
\newtheorem{Corollary}{Corollary}
\newcommand{\proof}[1]{{\bf Proof. }
        #1\begin{flushright}$\Box$\end{flushright}}

\baselineskip=20pt

\newfont{\elevenmib}{cmmib10 scaled\magstep1}
\newcommand{\preprint}{
   \begin{flushleft}
   \end{flushleft}\vspace{-1.3cm}
   \begin{flushright}\normalsize
   \end{flushright}}
\newcommand{\Title}[1]{{\baselineskip=26pt
   \begin{center} \Large \bf #1 \\ \ \\ \end{center}}}
\newcommand{\Author}{\begin{center}
   \large \bf
Junpeng Cao${}^{a, b}$,Wen-Li Yang${}^{c,d}
\footnote{Corresponding author: wlyang@nwu.edu.cn}$, Kangjie
Shi${}^c$ and~Yupeng Wang${}^{a,b}\footnote{Corresponding
author: yupeng@iphy.ac.cn}$
 \end{center}}
\newcommand{\Address}{\begin{center}

     ${}^a$Beijing National Laboratory for Condensed Matter
           Physics, Institute of Physics, Chinese Academy of Sciences, Beijing
           100190, China\\
     ${}^b$Collaborative Innovation Center of Quantum Matter, Beijing,
     China\\
     ${}^c$Institute of Modern Physics, Northwest University,
     Xian 710069, China \\
     ${}^d$Beijing Center for Mathematics and Information Interdisciplinary Sciences, Beijing, 100048,  China
   \end{center}}
\newcommand{\Accepted}[1]{\begin{center}
   {\large \sf #1}\\ \vspace{1mm}{\small \sf Accepted for Publication}
   \end{center}}

\preprint \thispagestyle{empty}
\bigskip\bigskip\bigskip

\Title{Exact solution of the alternating XXZ spin chain with generic non-diagonal boundaries } \vspace{1cm}
\Author

\Address \vspace{1cm}

\begin{abstract}

The integrable XXZ alternating spin chain with generic non-diagonal
boundary terms specified by  the most general non-diagonal
$K$-matrices is studied via the off-diagonal Bethe Ansatz method.
Based on the intrinsic properties of the fused $R$-matrices  and
$K$-matrices, we obtain certain closed operator identities and
conditions, which allow us to construct an inhomogeneous $T-Q$
relation and the associated Bethe Ansatz equations accounting for
the eigenvalues of the transfer matrix.

\vspace{1truecm} \noindent {\it PACS:} 75.10.Pq, 02.30.Ik, 05.30.Jp

\end{abstract}
\newpage

\section{Introduction}
\label{intro} \setcounter{equation}{0} There has been significant
focus of efforts on solving integrable quantum spin chains for many
years due to their numerous and  still growing applications in
string and super-symmetric Yang-Mills theories
\cite{Mad98,Dol03,Che04,Bei12}, statistical physics \cite{bax,McC10},
low-dimensional condensed matter physics \cite{kor2}, and even
some mathematical areas such as quantum groups \cite{Cha94}. Among
them, the XXZ spin chain (with various spins) plays a fundamental or
guiding role \cite{Kor93}. The Bethe Ansatz solution \cite{Bet31} of the
spin-$\frac{1}{2}$ XXZ chain  with periodic boundary condition (or
closed chain) was first given by Orbach \cite{oba} and revisited by
Yang and Yang \cite{Yan66} and many others.   The $s=1$ integrable
spin chain was first proposed by Zamalodchikov and Fateev \cite{zf}.
Its generalization to arbitrary $s$ cases was subsequently
constructed via the fusion techniques \cite{fusion,fusion-1} based on the
fundamental $s=\frac12$ representations of the Yang-Baxter equation
\cite{yang,bax}, an important equation which eventually led to the
discovery of the Quantum Inverse Scattering Method \cite{Skl78} and
Quantum Groups \cite{Cha94,fusion,fusion-1,Dri86,Jim85}. For the open spin-$\frac{1}{2}$
chain \cite{Gau71,Alc87}, Sklyanin \cite{Sk} proposed a systematic
method to construct and to diagonalize a commuting transfer matrix,
based on solutions of the boundary Yang-Baxter equation or the
reflection equation (RE) \cite{Che84}. Since then it directly
stimulated a great deal of studies on the exact solutions of the
quantum integrable models with open boundaries. A striking feature
of the reflection equation is that it allows non-diagonal solutions
\cite{de1,GZ}, which breaks the usual $U(1)$-symmetry (i.e., the
total spin is not anymore conserved) and leads to the corresponding
eigenvalue problem quite frustrated. Many efforts had been made
\cite{CLSW,Nep04,Yan04-1,Gie05,Gie05-1,Doi06,Yan06,Baj06,YZ3,Bas07,Gal08,Fra08,Nic12,bell}
to approach this nontrivial problem. However, in a long period of
time, the Bethe Ansatz solutions could only be obtained for either
constrained boundary parameters \cite{CLSW} or special crossing
parameters \cite{Nep04} associated with spin-$\frac12$ chains or
with spin-$s$ chains \cite{Fra07,mur,barmur,matins}.

Recently, a method for solving the eigenvalue problem of integrable
models with generic boundary conditions, i.e., the off-diagonal
Bethe Ansatz (ODBA) method was proposed in \cite{cao1} (\cite{Wan15} for the details) and then several
long-standing models \cite{cao1,lcysw14,ZCYSW14}  were solved.
Subsequently, its applications to integrable models beyond $A$-type
\cite{Hao14} and to the high spin XXX open chain \cite{Cao14-1} were
performed, and the nested-version of ODBA for the models associated
with $su(n)$ algebra \cite{Cao14} was developed. In addition, the
method for thermodynamic analysis  based on the ODBA solutions
\cite{Li14} was also proposed. It should be noted that two other
promising methods, namely, the $q$-Onsager algebra method
\cite{Bas07} and the separation of variables method \cite{nicc14-1}
were also  used to approach the spin-$\frac12$ chains with generic
integrable boundaries. Very recently based on the inhomogeneous $T-Q$ relations \cite{cao1} obtained by
ODBA,  the corresponding Bethe states, which have well-defined homogeneous limits, have been
proposed for  the open chains  by the modified algebraic Bethe ansatz \cite{sigma,Bel14} and
by ODBA \cite{Cao14-2, Zhan15}.

The high spin chains with periodic and  diagonal boundaries have
been extensively studied in the literature
\cite{zf,fusion,MNR,FLSU,Doi02}. So far the Bethe Ansatz solutions
of the models with non-diagonal boundaries were known only for some
special cases such as the boundary parameters obeying some
constraint \cite{Fra07} or  the crossing parameter (or anisotropy
constant) $\eta$  taking some special value (e.g., roots of unity)
\cite{mur,barmur,Cao14-1}. Moreover, the XXZ chain can be
generalized to integrable alternating spin chain
\cite{Doi02,DeV92,Ala93,DeV94,Doe96}, i.e., an inhomogeneous chain
with spin $s$ at odd sites and spin $s'$ at even sites. The
alternating spin chains have many applications in lower dimensional
quantum field theories such as the $SU(2)$ principle Chiral model
\cite{Doi01,Sha78} and the super-symmetric sine-Gordon model
\cite{Ina95, Baj02}. In this paper we shall investigate the Bethe Ansatz
solution of the integrable  XXZ alternating spin chain with an
arbitrary  $\eta$ and  generic non-diagonal boundary terms specified
by  the most general non-diagonal $K$-matrices via the ODBA.

The outline of the paper is  as follows. Section 2 serves as an
introduction to our notations and some basic ingredients. After
briefly reviewing the fusion procedures for the $R$-matrix
\cite{fusion,Jim85,Kul81} and the associated $K$-matrices  from the
fundamental spin-$\frac{1}{2}$ ones \cite{MNR,fusion2,Zho96}, we introduce
the corresponding transfer matrix of the open XXZ alternating spin
chain with the most generic non-diagonal boundary $K$-matrices and
the fused transfer matrices following the method in \cite{Doi02}. In
Section 3, using the method in \cite{Fra07} we derive the
corresponding fusion hierarchy  of the high spin fused transfer
matrices and give certain closed  operator product identities for
the fundamental transfer matrix by using some intrinsic properties
of the high spin  $R$-matrix ($R^{(l,l)}(u)$) and $K$-matrices
($K^{\pm(\frac{1}{2})}(u)$). The asymptotic behavior of the transfer
matrix is also obtained. The resulting conditions completely
characterize the eigenvalues of the fundamental transfer matrix (as
a consequence, also determine the eigenvalues of all the high spin
fused transfer matrices). Section 4 is devoted to the construction
of the inhomogeneous $T-Q$ relation and the corresponding Bethe
Ansatz equations (BAEs). In section 5, we summarize our results and
give some discussions. Some detailed technical proofs are given in
Appendix A$\&$B.

\section{Transfer matrices}
\label{aniso} \setcounter{equation}{0}

\subsection{Fusion of the $R$-matrices and the $K$-matrices}
Let us fixed a generic complex $\eta$ and two positive numbers $s,
s' \in \{\frac{1}{2}, 1, \frac{3}{2}, \ldots\}$. Throughout, $V_i$
denotes  a $(2l_i+1)$-dimensional linear space ($\Cb^{2l_i+1}$)
which endows an irreducible representation of the quantum algebra
$\U$ \cite{Cha94} with spin $l_i$, where $q=e^{\eta}$. The
definition of $\U$ and its spin-$l$ representation are given in
Appendix A. The $R$-matrix  $R^{(l_i,l_j)}_{ij}(u)$, denoted as the
spin-$(l_i,l_j)$ $R$-matrix, is a linear operator acting in
$V_i\otimes V_j$. The $R$-matrix satisfies the following quantum
Yang-Baxter equation (QYBE) \cite{yang,bax} \bea
R_{12}^{(l_1,l_2)}(u-v) R_{13}^{(l_1,l_3)}(u)R_{23}^{(l_2,l_3)}(v)=
R_{23}^{(l_2,l_3)}(v)
R_{13}^{(l_1,l_3)}(u)R_{12}^{(l_1,l_2)}(u-v).\label{QYBE} \eea Here
and below we adopt the standard notations: for any matrix $A\in {\rm
End}({\rm V})$, $A_j$ is an embedding operator in the tensor space
${ V}\otimes { V}\otimes\cdots$, which acts as $A$ on the $j$-th
space and as identity on the other factor spaces; $R_{ij}(u)$ is an
embedding operator of $R$-matrix in the tensor space, which acts as
identity on the factor spaces except for the $i$-th and $j$-th ones.

The fundamental spin-$(\frac{1}{2},l)$ $R$-matrix
$R^{(\frac{1}{2},l)}_{12}(u)$ (also called the $L$-operator
\cite{Kor93}) defined in spin-$\frac{1}{2}$ (i.e., two-dimensional)
auxiliary space and spin-$l$ (i.e., $(2l+1)$-dimensional) quantum
space is given by \cite{fusion,fusion-1} \bea
R^{(\frac{1}{2},l)}_{12}(u)&=&\sinh\lt(u+\eta(\frac{1}{2}+\s_1^3\,S_2^3)\rt)
+\sinh\eta\lt(\s_1^+\,S_2^-+\s_1^-\,S_2^+\rt)\no\\[6pt]
&=&\lt(\begin{array}{cc}\sinh(u+\frac{\eta}{2}+\eta S^3_2)&\sinh\eta \,\,S^-_2\\[2pt]
\sinh\eta\,\, S^+_2&\sinh(u+\frac{\eta}{2}-\eta
S^3_2)\end{array}\rt),\label{R-1s} \eea where $\eta$ is the
so-called crossing parameter, ${ \s^3,\,\s^{\pm}}$ are the Pauli
matrices and ${S^3,\,S^{\pm}}$ are the spin-$l$ realizations given
by (\ref{Rep-1})-(\ref{Rep-3}) of the generators of the quantum
algebra $\U$.   For the simplest case, i.e., $l=\frac{1}{2}$  the
corresponding $R$-matrix reads \bea R^{(\frac{1}{2},\frac{1}{2})}(u)
= \left(
\begin{array}{cccc}
    \sinh(u + \eta) &0            &0           &0            \\
    0                 &\sinh u     & \sinh\eta  &0            \\
    0                 &\sinh\eta   &  \sinh u    &0            \\
    0                 &0            &0           &\sinh(u + \eta)
\end{array} \right).
\label{R-matrix11} \eea \noindent Besides the QYBE (\ref{QYBE}), the
above well-known trigonometric  six-vertex $R$-matrix  also enjoys
the following properties, \bea &&\mbox{ Initial
condition}:\,R^{(\frac{1}{2},\frac{1}{2})}_{12}(0)=\sinh\eta P_{12},\label{Int-R}\\[6pt]
&&\mbox{ Unitary
relation}:\,R^{(\frac{1}{2},\frac{1}{2})}_{12}(u)R^{(\frac{1}{2},\frac{1}{2})}_{21}(-u)=
-\xi(u)\,{\rm id}, \\[6pt]
&& \qquad \qquad\qquad \qquad \xi(u)=\sinh(u+\eta)\sinh(u-\eta),\label{Unitarity}\\[6pt]
&&\mbox{ Crossing
relation}:\,R^{(\frac{1}{2},\frac{1}{2})}_{12}(u)=V_1\{R^{(\frac{1}{2},\frac{1}{2})}_{12}(-u-\eta)\}^{t_1}V_1,\quad
V=-i\s^y,
\label{crosing-unitarity}\\[6pt]
&&\mbox{
PT-symmetry}:\,R^{(\frac{1}{2},\frac{1}{2})}_{12}(u)=R^{(\frac{1}{2},\frac{1}{2})}_{21}(u)
=\{R^{(\frac{1}{2},\frac{1}{2})}_{12}(u)\}^{t_1\,t_2},\label{PT}\\[6pt]
&&\mbox{Antisymmetric-fusion conditions}:\,R^{(\frac{1}{2},\frac{1}{2})}_{12}(-\eta)\propto\, P_{12}^{-}.\label{Fusion-Con-1}\\[6pt]
&&\mbox{Symmetric-fusion conditions}:
\no \\[6pt]
&& \qquad \qquad\qquad \qquad
R^{(\frac{1}{2},\frac{1}{2})}_{12}(\eta) \propto \,{\rm
Diag}(\cosh\eta,1,1,\cosh\eta)\,P_{12}^{+}.\label{Fusion-Con-2} \eea
Here
$R^{(\frac{1}{2},\frac{1}{2})}_{21}(u)=P_{12}R^{(\frac{1}{2},\frac{1}{2})}_{12}(u)P_{12}$
with $P_{12}$ being the permutation operator between the tensor
product space of the spin-$\frac{1}{2}$ vector spaces;
$P^{\pm}_{12}=\frac{1}{2}(1\pm P_{12})$;  and $t_i$ denotes
transposition in the $i$-th space. The property (\ref{Fusion-Con-2})
allows one to construct the spin-$(j, l)$ $R$-matrix by using the
symmetric fusion procedure \cite{fusion,fusion-1} \be  R^{(j,l)}_{\{1 \cdots
2j\} \bar 1}(u) &=& B_{1\ldots 2j}A_{1\ldots 2j}P^{+}_{1 \cdots
2j} \prod_{k=1}^{2j}\lt\{R^{(\frac{1}{2},l)}_{k,\,\bar
1}(u+(k-j-\frac{1}{2})\eta)\rt\}\no \\[6pt]
&& \times
 P^{+}_{1 \cdots 2j}A^{-1}_{1\ldots 2j}B^{-1}_{1\ldots 2j}, \label{fused-Rjl}
\ee where $P^{+}_{1 \cdots 2j}$ is the symmetric projector given
by \bea
P^{+}_{1,\cdots,2j}=\frac{1}{(2j)!}\prod_{k=1}^{2j}\lt(\sum_{l=1}^k
P_{l\,k}\rt),\label{Symmetric-P} \eea and the $u$-independent
matrices $B_{1\ldots 2j}$ and $A_{1\ldots 2j}$ are given in
\cite{Nep04}. It is remarked that the $R$-matrices in the products
(\ref{fused-Rjl})  are ordered in the order of increasing $k$ and
that  the fused $R$-matrices (\ref{R-1s}) and (\ref{fused-Rjl})
satisfy the associated QYBE (\ref{QYBE}). Direct calculation  shows
that the spin-$(l,l)$ $R$-matrix is  given by
(\ref{R-matrix-relation}) \cite{Jim85, Byt03}. In particular, the
fused spin-$(l,l)$ $R$-matrix satisfies  the following important
properties \bea && \mbox{ Initial
condition}:\,R^{(l,l)}_{12}(0)\propto\, \mathbf{P}_{12},\label{Int-R-ll}\\[6pt]
&& \mbox{ Fusion
condition}:\,R^{(l,l)}_{12}(-\eta)\propto\,\mathbf{P}_{12}^{(0)}(l),
\label{Fusion-Con-3} \eea where $\mathbf{P}$ is the permutation
operator between the tensor product space of the spin-$l$ vector
spaces,
 and the projector $\mathbf{P}^{(0)}(l)$ is related to the projector $\bar{\mathbf{P}}^{(0)}(l)$ given by (\ref{Singlet-1}).
Namely, the projector $\mathbf{P}^{(0)}(l)$ is given by \bea
{\mathbf{P}}^{(0)}(l)=|\Phi^{(l)}_0\rangle\langle\Phi^{(l)}_0|,\quad
|\Phi^{(l)}_0\rangle=\frac{1}{\sqrt{2l+1}}\sum_{k=0}^{2l}(-1)^k|l-k\rangle\otimes|-l+k\rangle,\label{Singlet}
\eea where $\{|m\rangle|m=l,l-1,\ldots,-l\}$ forms an orthonormal
basis of the spin-$l$ space. The very properties (\ref{Int-R-ll})
and (\ref{Fusion-Con-3}) are the analogs of (\ref{Int-R}) and
(\ref{Fusion-Con-1}) for the higher spin case.

Having defined the fused-$R$ matrices, one can analogously construct
the fused-$K$ matrices by using the methods developed in
\cite{MNR,fusion2,Zho96} as follows.  The fused $K^-$ matrices (e.g
the spin-$j$ $K^-$ matrix) is given by \be K^{- (j)}_{\{a\}}(u) &=&
B_{a_1\ldots a_{2j}}A_{a_1\ldots a_{2j}} P_{a_1\ldots a_{2j}}^{+} \prod_{k=1}^{2j} \Bigg\{
\left[ \prod_{l=1}^{k-1} R^{(\frac{1}{2},\frac{1}{2})}_{a_{l}a_{k}}
(2u+(k+l-2j-1)\eta) \right] \non \\[6pt]
&&\times  K^{- (\frac{1}{2})}_{a_{k}}(u+(k-j-\frac{1}{2})\eta)
\Bigg\} P_{a_1\ldots a_{2j}}^{+} \,A^{-1}_{a_1\ldots a_{2j}}B^{-1}_{a_1\ldots a_{2j}},
\label{fusedKmatrix0103} \ee where
$R^{(\frac{1}{2},\frac{1}{2})}(u)$ is given by (\ref{R-matrix11})
and the $u$-independent matrices $B_{a_1\ldots a_{2j}}$ and $A_{a_1\ldots
a_{2j}}$ are given in \cite{Nep04}. In this paper we consider  the most
general non-diagonal spin-$\frac{1}{2}$ $K^-$ matrix $K^{-
(\frac{1}{2})}(u)$ given by \cite{de1, GZ} \bea
K^{-(\frac{1}{2})}(u)&=&\lt(\begin{array}{ll}K^-_{11}(u)&K^-_{12}(u)\\[6pt]
K^-_{21}(u)&K^-_{22}(u)\end{array}\rt),\no\\[6pt]
K^-_{11}(u)&=&2\lt(\sinh(\a_-)\cosh(\b_{-})\cosh(u)
+\cosh(\a_-)\sinh(\b_-)\sinh(u)\rt),\no\\[6pt]
K^-_{22}(u)&=&2\lt(\sinh(\a_-)\cosh(\b_{-})\cosh(u)
-\cosh(\a_-)\sinh(\b_-)\sinh(u)\rt),\no\\[6pt]
K^-_{12}(u)&=&e^{\theta_-}\sinh(2u),\quad
K^-_{21}(u)=e^{-\theta_-}\sinh(2u),\label{K-}\eea where
$\a_{-},\,\b_{-},\,\theta_{-}$ are some boundary parameters. It is
noted that the products  of braces $\{\ldots\}$ in the above
equation are ordered in the order of increasing $k$. The fused $K^{-
(j)}_{\{a\}}(u)$ matrices satisfy the following reflection equation
\cite{Che84,Fra07} \be \lefteqn{R^{(j,s)}_{\{a\} \{b\}}(u-v)\, K^{-
(j)}_{\{a\}}(u)\,
R^{(s,j)}_{\{b\} \{a\}}(u+v)\, K^{- (s)}_{\{b\}}(v)}\non \\[6pt]
& & =K^{- (s)}_{\{b\}}(v)\, R^{(j,s)}_{\{a\} \{b\}}(u+v)\, K^{-
(j)}_{\{a\}}(u)\, R^{(s,j)}_{\{b\} \{a\}}(u-v) \,.\label{RE} \ee The
fused dual reflection matrices $K^{+ (j) }$ \cite{Sk} are given by
\be K^{+ (j)}_{\{a\}}(u) = {1\over f^{(j)}(u)}\,K^{- (j)}_{\{a\}}
(-u-\eta)\Big\vert_{(\a_-,\b_-,\theta_-)\rightarrow
(-\a_+,-\b_+,\theta_+)} \,, \label{Correspond}\ee where the
normalization factor $f^{(j)}(u)$ is, \be f^{(j)}(u) =
\prod_{l=1}^{2j-1}\prod_{k=1}^{l} [-\xi( 2u + (l+k+1-2j)\eta) ],
\no \ee with the function $\xi(u)$ given
by (\ref{Unitarity}). Particularly, the fundamental one
$K^{+(\frac{1}{2})}(u)$ is
\begin{eqnarray}
K^{+(\frac12)}(u)=\lt.K^{-
(\frac12)}(-u-\eta)\rt|_{(\a_-,\b_-,\theta_-)\rightarrow
(-\a_+,-\b_+,\theta_+)},\ \label{K-0102}
\end{eqnarray}
where $\a_{+},\,\b_{+},\,\theta_{+}$ are some other boundary
parameters.

\subsection{Open alternating spin chains and its fused ones}
Periodic alternating spin chains were first studied in \cite{DeV92,
Ala93, DeV94, Doe96, Doi01} and then generalized to the open chain
case \cite{Sk, Kul91, Doi02}. Following \cite{Sk}, one can construct
the associated transfer matrix for an alternating XXZ spin chain
\cite{Doi04}, namely, an inhomogeneous chain with spin $s$ at odd
sites and spin $s'$ at even sites. Let us denote the transfer matrix
$t^{(j,(s,s'))}(u)$ whose auxiliary space is spin-$j$ ($(2j + 1)$-
dimensional) and each of its $2N$ quantum spaces with  alternative
spins, for any $j, s, s' \in \{\frac{1}{2}, 1, \frac{3}{2},
\ldots\}$. The fused (or the spin-$(j,(s,s'))$) transfer matrix
$t^{(j,(s,s'))}(u)$ can be constructed by the fused $R$-matrices and
$K$-matrices as follows \cite{Sk,Fra07}
\be t^{(j,(s,s'))}(u) =
\tr_{\{a\}} K^{+ (j)}_{\{a\}}(u)\, T^{(j,(s,s'))}_{\{a\}}(u)\, K^{-
(j)}_{\{a\}}(u)\, \hat T^{(j,(s,s'))}_{\{a\}}(u) \,,
\label{Fused-transfer} \ee where $T^{(j,(s,s'))}_{\{a\}}(u)$ and
$\hat T^{(j,(s,s'))}_{\{a\}}(u)$ are the fused one-row monodromy
matrices given by \bea T^{(j,(s,s'))}_{\{a\}}(u) &=&
R^{(j,s')}_{\{a\}, 2N}(u-\theta_{2N})R^{(j,s)}_{\{a\},
2N-1}(u-\theta_{2N-1}) \ldots \no \\[6pt]
&&\times R^{(j,s')}_{\{a\}, 2}(u-\theta_2)
R^{(j,s)}_{\{a\}, 1}(u-\theta_1) \,, \no \\[6pt]
\hat T^{(j,(s,s'))}_{\{a\}}(u) &=&
R^{(s,j)}_{1,\{a\}}(u+\theta_1)R^{(s',j)}_{2,\{a\}}(u+\theta_2)
\ldots \no \\[6pt]
&&\times R^{(s,j)}_{2N-1,\{a\}}(u+\theta_{2N-1})
R^{(s',j)}_{2N,\{a\}}(u+\theta_{2N}).\no \ee Here
$\{\theta_j|j=1,\ldots, 2N\}$ are arbitrary free complex parameters
which are usually called the inhomogeneous parameters. The QYBE
(\ref{QYBE}), the reflection equation (\ref{RE}) and its dual one
which can be deduced by the correspondence (\ref{Correspond})
between the $K^-$-matrices and $K^+$-matrices leads to \cite{Sk}
that these transfer matrices commute for different values of
spectral parameter, any $j , j' \in \{\frac{1}{2}, 1, \frac{3}{2},
\ldots \}$ and any $s ,s'\in \{\frac{1}{2}, 1, \frac{3}{2}, \ldots
\}$, \be \left[ t^{(j,(s,s'))}(u) \,, t^{(j',(s,s'))}(v) \right] = 0
\,. \label{commutativity0103} \ee Therefore $\{t^{(j,(s,s'))}\}$
serve as the generating functionals of the conserved quantities of
the associated model, whose Hamiltonian can be given by some derivative of the logarim of the associated
transfer matrix $t^{(j,(s,s'))}(u)$ at some special points \cite{Doi02,Byt03,Doi04},  and thus ensure the integrability of the model.

\section{Fusion hierarchy and the operator identities}
\label{anisoopen}\setcounter{equation}{0}

\subsection{Operator identities}
In the following part of the paper,  let us denote the fused
transfer matrices $\{t^{(j,(s,s'))}(u)\}$ given by
(\ref{Fused-transfer}) by $\{t^{(j)}(u)\}$ for simplicity. One  may
verify that these fused transfer matrices obey the following fusion
hierarchy relation following the method in \cite{MNR,fusion2,Fra07}
\be t^{(\frac{1}{2})}(u)\, t^{(j-\frac{1}{2})}(u- j\eta)&=&
t^{(j)}(u-(j-\frac{1}{2})\eta) + \delta^{(s,s')}(u)\,
t^{(j-1)}(u-(j+\frac{1}{2})\eta),\no\\[6pt]
&& j =\frac 12, 1, \frac{3}{2}, \cdots, \label{hierarchy0103} \ee
where we have used the convention $t^{(0)}={\rm id}$. The
coefficient function $\delta^{(s,s')}(u)$, which is related to the
quantum determinant, is given by \be
\hspace{-1.5cm}\delta^{(s,s')}(u) &=& 2^4\frac{\sinh(2u-2\eta)
\sinh(2u+2\eta)}{\sinh(2u+\eta)\sinh(2u-\eta)}\sinh(u+\a_-)\sinh(u-\a_-)
\cosh(u+\b_-)\no\\[6pt]
&& \times \cosh(u-\b_-)\sinh(u+\a_+)\sinh(u-\a_+)
\cosh(u+\b_+)\cosh(u-\b_+)\no\\[6pt]
&&  \times \prod_{l=1}^N \sinh(u-\theta_{2l-1}+(\frac{1}{2}+s)\eta)\sinh(u+\theta_{2l-1}+(\frac{1}{2}+s)\eta)\no\\[6pt]
&&  \times \prod_{l=1}^N \sinh(u-\theta_{2l}+(\frac{1}{2}+s')\eta)\sinh(u+\theta_{2l}+(\frac{1}{2}+s')\eta)\no\\[6pt]
&&  \times \prod_{l=1}^N \sinh(u-\theta_{2l-1}-(\frac{1}{2}+s)\eta)\sinh(u+\theta_{2l-1}-(\frac{1}{2}+s)\eta)\no\\[6pt]
&&  \times \prod_{l=1}^N
\sinh(u-\theta_{2l}-(\frac{1}{2}+s')\eta)\sinh(u+\theta_{2l}-(\frac{1}{2}+s')\eta).\label{delta0103}
\ee Using the recursive relation (\ref{hierarchy0103}), we can
express the fused transfer matrix $t^{(j)}(u)$ in terms of the
fundamental one $t^{(\frac{1}{2})}(u)$ with a $2j$-order operator product
relation as follows: \bea
 \hspace{-1cm}t^{(j)}(u)&=&t^{(\frac{1}{2})}(u+(j-\frac{1}{2})\eta)\,t^{(\frac{1}{2})}(u+(j-\frac{1}{2})\eta-\eta)\ldots
                t^{(\frac{1}{2})}(u-(j-\frac{1}{2})\eta)\no\\[6pt]
             &&-\d^{(s,s')}(u+(j-\frac{1}{2})\eta)\,t^{(\frac{1}{2})}(u+(j-\frac{1}{2})\eta-2\eta)\ldots
             \,t^{(\frac{1}{2})}(u-(j-\frac{1}{2})\eta)\no\\[6pt]
             &&-\d^{(s,s')}(u+(j-\frac{1}{2})\eta-\eta)\,t^{(\frac{1}{2})}(u+(j-\frac{1}{2})\eta)\,
             t^{(\frac{1}{2})}(u+(j-\frac{1}{2})\eta-3\eta)\no\\[6pt]
             &&\quad\quad \times\ldots
             \,t^{(\frac{1}{2})}(u-(j-\frac{1}{2})\eta)\no\\[6pt]
             &&\vdots\no\\
             &&-\d^{(s,s')}(u\hspace{-0.04truecm}-\hspace{-0.04truecm}(j\hspace{-0.04truecm}-\hspace{-0.04truecm}\frac{1}{2})\eta
             \hspace{-0.04truecm}+\hspace{-0.04truecm}\eta)\,t^{(\frac{1}{2})}
             (u\hspace{-0.04truecm}+\hspace{-0.04truecm}(j\hspace{-0.04truecm}-\hspace{-0.04truecm}\frac{1}{2})\eta)\ldots
             \,t^{(\frac{1}{2})}(u\hspace{-0.04truecm}-\hspace{-0.04truecm}
             (j\hspace{-0.04truecm}-\hspace{-0.04truecm}\frac{1}{2})\eta\hspace{-0.04truecm}+\hspace{-0.04truecm}2\eta)\no\\[6pt]
             &&+\ldots.\label{Fusion-Hier}
\eea For examples, the first three fused transfer matrices are given
by \bea
  t^{(1)}(u)&=& t^{(\frac{1}{2})}(u+\frac{\eta}{2})\,t^{(\frac{1}{2})}(u-\frac{\eta}{2})-\d^{(s,s')}(u+\frac{\eta}{2}),\label{Fused-T-1}\\[6pt]
  t^{(\frac{3}{2})}(u)&=& t^{(\frac{1}{2})}(u+\eta)\,t^{(\frac{1}{2})}(u)\,t^{(\frac{1}{2})}(u-\eta)
                           -\d^{(s,s')}(u+\eta)\,t^{(\frac{1}{2})}(u-\eta)\no\\[6pt]
                        && -\d^{(s,s')}(u)\,t^{(\frac{1}{2})}(u+\eta),\label{Fused-T-2}\\[6pt]
  t^{(2)}(u)&=& t^{(\frac{1}{2})}(u+\frac{3\eta}{2})\,t^{(\frac{1}{2})}(u+\frac{\eta}{2})
                   \,t^{(\frac{1}{2})}(u-\frac{\eta}{2})\,t^{(\frac{1}{2})}(u-\frac{3\eta}{2})\no\\[6pt]
              &&-\d^{(s,s')}(u+\frac{3\eta}{2})\,t^{(\frac{1}{2})}(u-\frac{\eta}{2})\,t^{(\frac{1}{2})}(u-\frac{3\eta}{2})\no\\[6pt]
              &&-\d^{(s,s')}(u+\frac{\eta}{2})\,t^{(\frac{1}{2})}(u+\frac{3\eta}{2})\,t^{(\frac{1}{2})}(u-\frac{3\eta}{2})\no\\[6pt]
              &&-\d^{(s,s')}(u-\frac{\eta}{2})\,t^{(\frac{1}{2})}(u+\frac{3\eta}{2})\,t^{(\frac{1}{2})}(u+\frac{\eta}{2})\no\\[6pt]
              &&+\d^{(s,s')}(u+\frac{3\eta}{2})\,\d^{(s,s')}(u-\frac{\eta}{2}).\label{Fused-T-3}
\eea Keeping the very properties (\ref{Int-R-ll}) and
(\ref{Fusion-Con-3}) in mind and following the method developed  in
\cite{Cao14,Hao14},  we can derive the following relations with the help of QYBE (\ref{QYBE}),
\bea
\hspace{-0.8truecm}T^{(s)}_1(\theta_{2j-\hspace{-0.02truecm}1})\,
T^{(s)}_2(\theta_{2j-1}\hspace{-0.04truecm}-\hspace{-0.04truecm}\eta)\hspace{-0.2truecm}&=&\hspace{-0.2truecm}
\mathbf{P}_{21}^{(0)}(s)\,T^{(s)}_1(\theta_{2j-1})\,T^{(s)}_2(\theta_{2j-1}-\eta),\, j=1,\cdots,N. \\[6pt]
\hspace{-0.8truecm}\hat{T}^{(s)}_1(-\theta_{2j-1})\,\hat{T}^{(s)}_2(-\theta_{2j-1}-\eta)\hspace{-0.2truecm}&=&\hspace{-0.2truecm}
\mathbf{P}_{21}^{(0)}(s)\,\hat{T}^{(s)}_1(-\theta_{2j-1})\,\hat{T}^{(s)}_2(-\theta_{2j-1}-\eta),\, j=1,\cdots,N,
\eea where we use $T^{(s)}(u)$ (or $\hat{T}^{(s)}(u)$) to denote the one-row monodromy matrix $T^{(s,(s,s'))}(u)$
(or $\hat{T}^{(s,(s,s'))}(u)$). The above relations and  the RE (\ref{RE}) enable us to derive the following relations of
the double-row monodromy matrix $\mathbb{T}^{(s)}(u)=T^{(s)}(u){K^-}^{(s)}(u)\hat{T}^{(s)}(u)$
\bea
&&\mathbb{T}^{(s)}_1(\theta_{2j-1})\,R^{(s,s)}_{2,1}(2\theta_{2j-1}-\eta)\,\mathbb{T}^{(s)}_2(\theta_{2j-1}-\eta)\no\\[6pt]
&&\quad\quad\quad\quad=\mathbf{P}_{21}^{(0)}(s)
\mathbb{T}^{(s)}_1(\theta_{2j-1})\,R^{(s,s)}_{2,1}(2\theta_{2j-1}-\eta)\,\mathbb{T}^{(s)}_2(\theta_{2j-1}-\eta)
,\, j=1,\cdots,N.
\eea Then the above relation and the dual of RE (which can be derived by the RE (\ref{RE}) and the
correspondence (\ref{Correspond})) give rise to that
the spin-$s$ transfer matrix satisfy the following
operator identities, \bea
&&t^{(s)}(\theta_{2j-1})\,\,t^{(s)}(\theta_{2j-1}-\eta)=\Delta^{(s)}(u)\lt|_{u=\theta_{2j-1}}\rt.\times{\rm id},\quad j=1,\ldots,N,\label{Operator-identity-1} \eea where the
function $\Delta^{(l)}(u)$ is \bea
\Delta^{(l)}(u)=\prod_{k=0}^{2l-1}\d^{(s,s')}(u-(l-\frac{1}{2})\eta+k\eta),\label{Deter-s}
\eea and the function $\d^{(s,s')}(u)$ is given by
(\ref{delta0103}). Using the similar method, we can derive  that the spin-$s'$ transfer matrix satisfy the following
operator identities, \bea
t^{(s')}(\theta_{2j})\,\,t^{(s')}(\theta_{2j}-\eta)=\Delta^{(s')}(u)\lt|_{u=\theta_{2j}}\rt.\times{\rm
id},\quad j=1,\ldots,N,\label{Operator-identity-2} \eea
where the function $\Delta^{(s')}(u)$ is given by (\ref{Deter-s}).

Now let us derive some properties of the fundamental transfer matrix
$t^{(\frac{1}{2})}(u)$. For this purpose we first list some
properties of the fundamental spin-($\frac{1}{2},l$) $R$-matrix
(\ref{R-1s}) which are some analogs of (\ref{Unitarity}) and
(\ref{crosing-unitarity}) for a generic $l$. With the help of the
commutation relations (\ref{Commututing-relations}) of $\U$ and
(\ref{Casi-2}), we can derive  the following generalized unitary relation
(c.f., (\ref{Unitarity})) \bea
R^{(\frac{1}{2},l)}_{12}(u)R^{(l,\frac{1}{2})}_{21}(-u)=
-\sinh(u+(\frac{1}{2}+l)\eta) \sinh(u-(\frac{1}{2}+l)\eta)\times
{\rm id}. \label{Unitarity-1} \eea Direct calculation shows that the
$R$-matrix $R^{(\frac{1}{2},l)}_{12}(u)$ also satisfies the
following properties \bea
&&R^{(\frac{1}{2},l)}_{12}(u)=V_1\{R^{(\frac{1}{2},l)}_{12}(-u-\eta)\}^{t_1}V_1,\quad V=-i\s^y,\label{crosing-unitarity-1}\\[6pt]
&& R^{(\frac{1}{2},l)}_{12}(u+i\pi)=-\s^z_1
R^{(\frac{1}{2},l)}_{12}(u)\s^z_1. \eea It is easy to check that the
fundamental spin-$\frac{1}{2}$ K-matrices $K^{\pm(\frac{1}{2})}(u)$
given by (\ref{K-}) and (\ref{K-0102}) enjoy  the following
properties \bea
&&K^{-(\frac{1}{2})}(0)=\frac{1}{2}tr(K^{-(\frac{1}{2})}(0))\times {\rm id},\label{K-1}\\[6pt]
&&K^{-(\frac{1}{2})}(\frac{i\pi}{2})=\frac{1}{2}tr(K^{-(\frac{1}{2})}(\frac{i\pi}{2})\s^z)\times
\s^z,\label{K-spcial}\\[6pt]
&&K^{\pm(\frac{1}{2})}(u+i\pi)=-\s^z\,K^{\pm(\frac{1}{2})}(u)\,\s^z.\label{K-3}
\eea The above relations, the explicit expressions of the
spin-$(\frac{1}{2},s)$ and spin-$(\frac{1}{2},s')$ $R$-matrices
given by (\ref{R-1s}) and the spin-$\frac{1}{2}$ $K$-matrices given
by (\ref{K-}) and (\ref{K-0102}) imply that the transfer matrix
$t^{(\frac{1}{2})}(u)$ satisfies the following properties: \bea
&&\hspace{-1cm}t^{(\frac{1}{2})}(u+i\pi)=t^{(\frac{1}{2})}(u),\label{Operator-perio}\\[6pt]
&&\hspace{-1cm}t^{(\frac{1}{2})}(-u-\eta)=t^{(\frac{1}{2})}(u),\label{Operator-crossing}\\[6pt]
&&\hspace{-1cm}t^{(\frac{1}{2})}(0)=-2^3\sinh\a_-\cosh\b_-\sinh\a_+\cosh\b_+\cosh\eta\,\no\\[6pt]
&&\hspace{-1cm}\quad\quad\quad\quad\times \prod_{l=1}^N\sinh(\theta_{2l-1}+(\frac{1}{2}+s)\eta)\sinh(-\theta_{2l-1}+(\frac{1}{2}+s)\eta)\no\\[6pt]
&&\hspace{-1cm}\quad\quad\quad\quad\times
\prod_{l=1}^N\sinh(\theta_{2l}+(\frac{1}{2}+s')\eta)\sinh(-\theta_{2l}+(\frac{1}{2}+s')\eta)\times
{\rm id},
 \label{Operator-Int-1}\\[6pt]
&&\hspace{-1cm}t^{(\frac{1}{2})}(\frac{i\pi}{2})=-2^3\cosh\a_-\sinh\b_-\cosh\a_+\sinh\b_+\cosh\eta\,\no\\[6pt]
&&\hspace{-1cm}\quad\quad\quad\quad\times\prod_{l=1}^N\sinh(\frac{i\pi}{2}+\theta_{2l-1}+(\frac{1}{2}+s)\eta)
\sinh(\frac{i\pi}{2}+\theta_{2l-1}-(\frac{1}{2}+s)\eta)\no\\[6pt]
&&\hspace{-1cm}\quad\quad\quad\quad\times\prod_{l=1}^N\sinh(\frac{i\pi}{2}+\theta_{2l}+(\frac{1}{2}+s')\eta)
\sinh(\frac{i\pi}{2}+\theta_{2l}-(\frac{1}{2}+s')\eta)\times {\rm
id},
 \label{Operator-Int-2}\\[6pt]
&&\hspace{-1cm}t^{(\frac{1}{2})}(u)\lt|_{u\rightarrow\infty}\rt.=
-\frac{\cosh(\theta_--\theta_+)e^{\pm[(4N+4)u+(2N+2)\eta]}}
{2^{4N+1}}\times {\rm id} +\ldots.\label{Operator-Asy} \eea

The analyticities of the spin-$(\frac{1}{2},s)$ and
spin-$(\frac{1}{2},s')$ $R$-matrices and spin-$\frac{1}{2}$
$K$-matrices and the property (\ref{Operator-Asy}) imply that the
fundamental transfer matrix $t^{(\frac{1}{2})}(u)$, as a function of
$u$, is a trignometric polynomial of degree $4N+4$. The fusion
hierarchy relation (\ref{hierarchy0103}) gives rise to that all the
other fused transfer matrix  $t^{(j)}(u)$ can be expressed in terms
of some sum of  products  of the fundamental one  (see
(\ref{Fusion-Hier})). Particularly, the spin-$s$ transfer matrix
$t^{(s)}(u)$ (or the spin-$s'$ transfer matrix $t^{(s')}(u)$) is
expressed in terms of the sum of products of $t^{(\frac{1}{2})}(u)$
with  orders up to $2s$ (or $2s'$). The very identities
(\ref{Operator-identity-1})-(\ref{Operator-identity-2}) then lead to
$2N$ constraints on the fundamental transfer matrix
$t^{(\frac{1}{2})}(u)$. Thus the relations
(\ref{Operator-identity-1})-(\ref{Operator-identity-2}) and
(\ref{Operator-perio})-(\ref{Operator-Asy}) completely  characterize
the eigenvalues of the fundamental transfer matrix
$t^{(\frac{1}{2})}(u)$ (as a consequence, also determine the
eigenvalues of all the transfer matrices $\{t^{(j)}(u)\}$).

\subsection{Functional relations of the eigenvalues}
The commutativity (\ref{commutativity0103}) of the fused transfer
matrices $\{t^{(j)}(u)\}$ with different spectral parameters implies
that they have common eigenstates. Let $|\Psi\rangle$  be a common
eigenstate of these fused transfer matrices, which does not depend
upon $u$, with the eigenvalue $\Lambda^{(j)}(u)$ , i.e., \bea
t^{(j)}(u)|\Psi\rangle =\Lambda^{(j)}(u)|\Psi\rangle. \no\eea The
fusion hierarchy relation (\ref{hierarchy0103}) of the fused
transfer matrices allows one to express all the eigenvalues
$\Lambda^{(j)}(u)$ in terms of the fundamental one
$\Lambda^{(\frac{1}{2})}(u)$ by the following recursive relations
\be \hspace{-1cm}\L^{(\frac{1}{2})}(u)\, \L^{(j-\frac{1}{2})} (u-
j\eta)&=&\hspace{-0.08truecm} \L^{(j)}(u- (j-\frac{1}{2})\eta) +
\delta^{(s,s')}(u)\,
\L^{(j-1)}(u-(j\hspace{-0.04truecm}+ \frac{1}{2})\eta),\no\\[6pt]
&& j =\frac{1}{2}, 1, \frac{3}{2}, \cdots. \label{Eigenvalue-hier}
\ee Here $\L^{(0)}(u)=1$ and the coefficient function
$\d^{(s,s')}(u)$ is given by (\ref{delta0103}). The very operator
identities (\ref{Operator-identity-1}) and
(\ref{Operator-identity-2}) of the fused transfer matrix at the
points $\theta_j$  imply that the eigenvalue $\Lambda^{()}(u)$
satisfies the similar relations \bea
&&\L^{(s)}(\theta_{2j-1})\,\,\L^{(s)}(\theta_{2j-1}-\eta)=\Delta^{(s)}(u)\lt|_{u=\theta_{2j-1}}\rt.,
\quad j=1,\ldots,N,\label{Eigenvalue-identity-1}\\[6pt]
&&\L^{(s')}(\theta_{2j})\,\,\L^{(s')}(\theta_{2j}-\eta)=\Delta^{(s')}(u)\lt|_{u=\theta_{2j}}\rt.
,\quad j=1,\ldots,N,\label{Eigenvalue-identity-2} \eea where the
function $\Delta^{(l)}(u)$ is given by (\ref{Deter-s}). The
properties of the transfer matrix $t^{(\frac{1}{2})}(u)$ given by
(\ref{Operator-perio})-(\ref{Operator-Asy}) give rise to that the
corresponding eigenvalue $\Lambda^{(\frac{1}{2})}(u)$ satisfies the
following relations \bea
&&\L^{(\frac{1}{2})}(u+i\pi)=\L^{(\frac{1}{2})}(u),\label{Eigenvalue-perio}\\[6pt]
&&\L^{(\frac{1}{2})}(-u-\eta)=\L^{(\frac{1}{2})}(u),\label{Eigenvalue-crossing}\\[6pt]
&&\L^{(\frac{1}{2})}(0)=-2^3\sinh\a_-\cosh\b_-\sinh\a_+\cosh\b_+\cosh\eta\,\no\\[6pt]
&&\quad\quad\quad\quad\times \prod_{l=1}^N\sinh(\theta_{2l-1}+(\frac{1}{2}+s)\eta)\sinh(-\theta_{2l-1}+(\frac{1}{2}+s)\eta)\no\\[6pt]
&&\quad\quad\quad\quad\times
\prod_{l=1}^N\sinh(\theta_{2l}+(\frac{1}{2}+s')\eta)\sinh(-\theta_{2l}+(\frac{1}{2}+s')\eta),
 \label{Eigenvalue-Int-1}\\[6pt]
&&\L^{(\frac{1}{2})}(\frac{i\pi}{2})=-2^3\cosh\a_-\sinh\b_-\cosh\a_+\sinh\b_+\cosh\eta\,\no\\[6pt]
&&\quad\quad\quad\quad\times\prod_{l=1}^N\sinh(\frac{i\pi}{2}+\theta_{2l-1}+(\frac{1}{2}+s)\eta)
\sinh(\frac{i\pi}{2}+\theta_{2l-1}-(\frac{1}{2}+s)\eta)\no\\[6pt]
&&\quad\quad\quad\quad\times\prod_{l=1}^N\sinh(\frac{i\pi}{2}+\theta_{2l}+(\frac{1}{2}+s')\eta)
\sinh(\frac{i\pi}{2}+\theta_{2l}-(\frac{1}{2}+s')\eta),
 \label{Eigenvalue-Int-2}\\[6pt]
&&\L^{(\frac{1}{2})}(u)\lt|_{u\rightarrow\pm\infty}\rt.=
-\frac{\cosh(\theta_--\theta_+)e^{\pm[(4N+4)u+(2N+2)\eta]}}
{2^{4N+1}} +\ldots.\label{Eigenvalue-Asy} \eea The analyticities of
the spin-$(\frac{1}{2},l)$ $R$-matrix and spin-$\frac{1}{2}$
$K$-matrices and the property (\ref{Eigenvalue-Asy}) imply that the
eigenvalue $\Lambda^{(\frac{1}{2})}(u)$ possesses the following
analytical property \bea \hspace{-2cm} \L^{(\frac{1}{2})}(u) \mbox{,
as a function of $u$, is a trigonometric polynomial of degree
$4N+4$}.\label{Eigenvalue-Anal} \eea Namely, $\L^{(\frac{1}{2})}(u)$
is a trigonometric polynomial of $u$ with $4N+5$ unknown
coefficients. The crossing relation (\ref{Eigenvalue-crossing})
reduces the number of the independent unknown coefficients to
$2N+3$. Therefore the  relations
(\ref{Eigenvalue-hier})-(\ref{Eigenvalue-Asy}) and the property (\ref{Eigenvalue-Anal}) completely
characterize  the spectrum of the fundamental transfer matrix
$t^{(\frac{1}{2})}(u)$.

\section{T-Q Ansatz and the associated BAEs} \label{ansatz}
\setcounter{equation}{0}
\subsection{Generic boundary parameters}
Following the method developed in \cite{cao1}, let us introduce the
following inhomogeneous $T-Q$ Ansatz for the eigenvalue
$\L^{(\frac{1}{2})}(u)$ of the fundamental transfer matrix
$t^{(\frac{1}{2})}(u)$ basing on the conditions
(\ref{Eigenvalue-hier})-(\ref{Eigenvalue-Anal}) \footnote{It was
shown in \cite{cao1} (see also \cite{Cao14-1}) that  there actually
exist a variety of apparent different $T-Q$ relations. However,
different forms of these $T-Q$ relations   only give different
parameterizations of the eigenvalues but not different states, and
each of them gives the complete set of the eigenvalues. For any
spins $s$, $s'$ and the total number of sites $2N$, one can always
choose a minimal number $2(s+s')N$ of the Bethe parameters to
parameterize $\L^{(\frac{1}{2})}(u)$ like (\ref{T-Q-ansatz-1}).},
\bea
 \L^{(\frac{1}{2})}(u)&=&a^{(s,s')}(u)\frac{Q(u-\eta)}{Q(u)}
                           +d^{(s,s')}(u)\frac{Q(u+\eta)}{Q(u)}\no\\[6pt]
 &&+2c\,\sinh 2u\,\sinh(2u+2\eta)\frac{F^{(s,s')}(u)}{Q(u)},\label{T-Q-ansatz-1}
\eea where  the functions $a^{(s,s')}(u)$, $d^{(s,s')}(u)$,
$F^{(s,s')}(u)$ and the constant $c$ are given by \bea
a^{(s,s')}(u)&=&-2^2\frac{\sinh(2u+2\eta)}{\sinh(2u+\eta)}\sinh(u-\a_-)\cosh(u-\b_-) \no\\[6pt]
&& \times \sinh(u-\a_+)\cosh(u-\b_+)\bar{A}^{(s,s')}(u),\label{a-function}\\[6pt]
&& \hspace{-2cm}d^{(s,s')}(u)=a^{(s,s')}(-u-\eta),\label{d-function}\\[6pt]
&& \hspace{-2cm}
\bar{A}^{(s,s')}(u)=\prod_{j=1}^N\sinh(u-\theta_{2j-1}+(\frac{1}{2}+s)\eta)\,
\sinh(u+\theta_{2j-1}+(\frac{1}{2}+s)\eta) \no \\[6pt]
&&  \times \prod_{j=1}^N\sinh(u-\theta_{2j}+(\frac{1}{2}+s')\eta)\,
\sinh(u+\theta_{2j}+(\frac{1}{2}+s')\eta),\label{A-function}\\[6pt]
F^{(s,s')}(u)&=&\prod_{j=1}^N\prod_{k=0}^{2s}\sinh(u-\theta_{2j-1}+(\frac{1}{2}-s+k)\eta)
\no\\[6pt]
&&\times \sinh(u+\theta_{2j-1}+(\frac{1}{2}-s+k)\eta)\no\\[6pt]
&&\times
\prod_{j=1}^N\prod_{k=0}^{2s'}\sinh(u-\theta_{2j}+(\frac{1}{2}-s'+k)\eta)
\no\\[6pt]
&&\times \sinh(u+\theta_{2j}+(\frac{1}{2}-s'+k)\eta),\label{F-function}\\[6pt]
&&\hspace{-2.2cm}c=\cosh(\a_-+\b_-+\a_++\b_++(1+2(s+s')N)\eta)-\cosh(\theta_--\theta_+).\label{c-constant}
\eea The  function $Q(u)$ is parameterized by $2(s+s')N$ parameters
$\{\l_j|j=1,\ldots,2(s+s')N\}$ as follow
\bea
Q(u)&=&\prod_{j=1}^{2(s+s')N}\sinh(u-\l_j)\,\sinh(u+\l_j+\eta)=Q(-u-\eta).\label{Q-function}
\eea From the explicit expressions
(\ref{T-Q-ansatz-1})-(\ref{Q-function}) of the Ansatz for
$\L^{(\frac{1}{2})}(u)$, one can easily  check that the $T-Q$ Ansatz
(\ref{T-Q-ansatz-1}) does satisfy the relations
(\ref{Eigenvalue-hier})-(\ref{Eigenvalue-Anal}) as follows. The
explicit expressions (\ref{a-function})-(\ref{F-function}) of the functions implies that
\bea
&&a^{(s,s')}(\theta_{2j-1}-(\frac{1}{2}+s)\eta)=0,\quad d^{(s,s')}(\theta_{2j-1}+(s-\frac{1}{2})\eta)=0,\no\\
&&F^{(s,s')}(\theta_{2j-1}+(s-\frac{1}{2}-k\eta))=0,\quad {\rm
for}\,\, k=0,1,\ldots,2s,\quad j=1,\ldots,N.\no \eea  Keeping the
above equations in mind  and iterating the fusion hierarchy relations
(\ref{Eigenvalue-hier}) from the fundamental one $\Lambda^{\frac{1}{2}}(u)$  given in terms of
the inhomogeneous $T$-$Q$ relation (\ref{T-Q-ansatz-1}), we can evaluate $\L^{(s)}(u)$ at the points
$\theta_{2j-1}$ and $\theta_{2j-1}-\eta$ as \bea
&&\hspace{-0.4truecm}\L^{(s)}(\theta_{2j-1})=
 \frac{Q(\theta_{2j-1}-
                    (s+\frac{1}{2})\eta)}
                    {Q(\theta_{2j-1}+
                    (s-\frac{1}{2})\eta)}
                    \prod_{k=0}^{2s-1}a^{(s,s')}
                    (\theta_{2j-1}+
                    (s-\frac{1}{2}-k)\eta),\\[6pt]
&& \hspace{-0.4truecm}\L^{(s)}(\theta_{2j-1}-\eta)=
                     \frac{Q(\theta_{2j-1}+
                     (s-\frac{1}{2})\eta)}
                     {Q(\theta_{2j-1}-
                     (s+\frac{1}{2})\eta)}
                    \prod_{k=0}^{2s-1}d^{(s,s')}(\theta_{2j-1}+
                    (s-\frac{3}{2}-k)\eta).
\eea The above equations yield \bea
\hspace{-0.4truecm}\L^{(s)}(\theta_{2j-1})\L^{(s)}
   (\theta_{2j-1}-\eta)
   &=& \prod_{k=0}^{2s-1}a^{(s,s')}(\theta_{2j-1}-
   (s-\frac{1}{2})\eta+k\eta) \no \\[6pt]
&&\times d^{(s,s')}(\theta_{2j-1}-
(s-\frac{1}{2})\eta+k\eta-\eta)\no\\[6pt]
&=& \prod_{k=0}^{2s-1}\d^{(s,s')}(\theta_{2j-1}-(s-\frac{1}{2})\eta+k\eta)\no\\[6pt]
&=& \Delta^{(s)}(\theta_{2j-1}),\quad j=1,\ldots,N. \eea In deriving
the second equality of the above equation, we have used the
following identity \bea \d^{(s,s')}(u)=a^{(s,s')}(u)\,
d^{(s,s')}(u-\eta). \eea Similarly, we can show that the anstaz
(\ref{T-Q-ansatz-1}) for $\L^{(\frac{1}{2})}(u)$ also satisfies the
relation (\ref{Eigenvalue-identity-2}). This means that the $T-Q$
Ansatz (\ref{T-Q-ansatz-1}) indeed satisfies the very functional
identities
(\ref{Eigenvalue-identity-1})-(\ref{Eigenvalue-identity-2}). From
the explicit expression (\ref{T-Q-ansatz-1}) one may find that the
$T-Q$ Ansatz might have some apparent simple poles at the following
points: \bea \l_j,\quad -\l_j-\eta,\quad
j=1,\ldots,2(s+s')N.\no \eea The regularity of the
transfer matrix implies that  the residues of the $T-Q$ Ansatz
(\ref{T-Q-ansatz-1}) at these points have to vanish which gives rise
to  the associated BAEs \bea && a^{(s,s)}(\l_j)Q(\l_j-\eta)
+d^{(s,s')}(\l_j)Q(\l_j+\eta) \no \\[6pt]
&&\quad  +2c\,\sinh 2\l_j\sinh(2\l_j+2\eta)F^{(s,s')}(\l_j)=0,
\no\\[6pt]
&&j=1,\ldots,2(s+s')N.\label{BAE-s-1} \eea

Finally we conclude that the $T-Q$ Ansatz (\ref{T-Q-ansatz-1})
indeed satisfies (\ref{Eigenvalue-hier})-(\ref{Eigenvalue-Anal})
provided that the $2(s+s')N$ parameters
$\{\l_j|j=1,\ldots,2(s+s')N\}$  satisfy the associated BAEs
(\ref{BAE-s-1}). Thus the $\L^{(\frac{1}{2})}(u)$ given by
(\ref{T-Q-ansatz-1}) becomes the eigenvalue of the transfer matrix
$t^{(\frac{1}{2})}(u)$ given by (\ref{Fused-transfer}) with
$j=\frac{1}{2}$. With the help of the recursive relation
(\ref{Eigenvalue-hier}), we can obtain the inhomogeneous $T-Q$
equations \footnote{It is noted that for the spin-$\frac{1}{2}$ open chain, such
equations can be obtained from a generating functions \cite{Nep13}.} for all the other $\L^{(j)}(u)$
from that of the
fundamental one $\L^{(\frac{1}{2})}(u)$.

\subsection{Constrained boundary parameters}

Following \cite{Yan06}, let us introduce four parameters
$\{\e_i|i=0,1,2,3\}$, each of which takes the values $\pm 1$. These
discrete parameters satisfy the relation \bea
\e_1\,\e_2\,\e_3=1.\label{Constraint-e} \eea
Similarly as that in previous subsection we can show that the eigenvalue $ \L^{(\frac{1}{2})}(u)$ of the fundamental transfer matrix
$t^{(\frac{1}{2})}(u)$ can also be parameterized
by any of the following inhomogeneous $T$-$Q$ relations,
\bea
 \L^{(\frac{1}{2})}(u)&=&a^{(s,s')}_{\pm}(u|\e_1,\e_2,\e_3)\frac{Q(u-\eta)}{Q(u)}
                           +d^{(s,s')}_{\pm}(u|\e_1,\e_2,\e_3)\frac{Q(u+\eta)}{Q(u)}\no\\[6pt]
 &&+2c_{\pm}(\e_1,\e_2,\e_3)\,\sinh 2u\,\sinh(2u+2\eta)\frac{F^{(s,s')}(u)}{Q(u)},\label{T-Q-ansatz-new-1}
\eea
where  the functions $F^{(s,s')}(u)$ and $Q(u)$ are given by (\ref{F-function}) and (\ref{Q-function}), and
the functions  $a_{\pm}^{(s,s')}(u)$ and $d_{\pm}^{(s,s')}(u)$ are  \bea
a_{\pm}^{(s,s')}(u|\e_1,\e_2,\e_3)&=&-2^2\e_2\,\frac{\sinh(2u+2\eta)}{\sinh(2u+\eta)}\sinh(u\mp\a_-)\cosh(u\mp\e_1\b_-) \no\\[6pt]
          &&  \times \sinh(u\mp\e_2\a_+)\cosh(u\mp\e_3\b_+)\bar{A}^{(s,s')}(u),\label{a-function-1}\\[6pt]
d_{\pm}^{(s,s')}(u|\e_1,\e_2,\e_3)&=&a_{\pm}^{(s,s')}(-u-\eta|\e_1,\e_2,\e_3).\label{d-function-1}
\eea The constant $c_{\pm}(\e_1,\e_2,\e_3)$ is
\bea
c_{\pm}=\e_2\,\cosh(\a_-+\e_1\b_-+\e_2\a_++\e_3\b_+\pm(1+2(s+s')N)\eta)-\cosh(\theta_--\theta_+).\label{c-constant-new-1}
\eea
The regularity of  each  $T-Q$ Ansatz
(\ref{T-Q-ansatz-new-1}) at the roots of $Q(u)$ gives rise
to  the resulting  BAEs which are similar as those of (\ref{BAE-s-1}). It is easy to check that
each $T-Q$ relation (\ref{T-Q-ansatz-new-1}) indeed satisfies (\ref{Eigenvalue-hier})-(\ref{Eigenvalue-Anal}).
The numerical analysis \cite{Nep13,Wan15} of the solutions of the associated BAEs for some small sites $N$ with $s=s'$ strongly suggests that
 each of the $T-Q$ relation (\ref{T-Q-ansatz-new-1}) may give a complete set of solutions of the transfer matrix.

In order to look for the solution of which the third term (i.e. inhomogeneous term) of (\ref{T-Q-ansatz-new-1}) vanishes, namely,
the solution of the form
\bea
 \L^{(\frac{1}{2})}(u)&=&a^{(s,s')}_{\pm}(u|\e_1,\e_2,\e_3)\frac{Q(u-\eta)}{Q(u)}
                           +d^{(s,s')}_{\pm}(u|\e_1,\e_2,\e_3)\frac{Q(u+\eta)}{Q(u)},\label{T-Q-H}
\eea
the asymptotic behavior (\ref{Eigenvalue-Asy})
gives rise to that  the 6 boundary
parameters $\a_{\pm}$, $\b_{\pm}$ and $\theta_{\pm}$ have to satisfy 
the corresponding constraint 
\bea
\a_-+\e_1\b_-+\e_2\a_++\e_3\b_++\e_0(\theta_--\theta_+)=k\eta+\frac{1-\e_2}{2}i\pi\quad
{\rm mod}(2i\pi),\label{Constraint-BP-1} \eea where $k$ is an integer
such that \bea 2(s+s')N-1\mp k=2M^{\pm},\quad
M^{\pm}=0,1,\ldots.\label{Constraint-BP-2} \eea
It is easy to check that
each $T-Q$ relation (\ref{T-Q-H}) indeed satisfies (\ref{Eigenvalue-hier})-(\ref{Eigenvalue-Int-2}) and (\ref{Eigenvalue-Anal}).
Moreover, the constraint (\ref{Constraint-BP-1})-(\ref{Constraint-BP-2}) makes (\ref{Eigenvalue-Asy}) satisfied.
This means that if the 6 boundary
parameters $\a_{\pm}$, $\b_{\pm}$ and $\theta_{\pm}$ satisfy any of the constraints (\ref{Constraint-BP-1})-(\ref{Constraint-BP-2}) the
inhomogeneous $T$-$Q$ relation (\ref{T-Q-ansatz-new-1}) reduces to two
conventional $T$-$Q$ relations \footnote{The solutions with the discrete parameters $\{\e_i|i=0,1,2,3\}$ was first
given in \cite{Yan06} and then discussed in \cite{nicc14-1}. }
\bea
 \L^{(\frac{1}{2})}_{\pm}(u)&=&a_{\pm}^{(s,s')}(u|\e_1,\e_2,\e_3)\frac{Q^{(\pm)}(u-\eta)}{Q^{(\pm)}(u)}
                           \no \\[6pt]
&&
+d_{\pm}^{(s,s')}(u|\e_1,\e_2,\e_3)\frac{Q^{(\pm)}(u+\eta)}{Q^{(\pm)}(u)},\label{T-Q-ansatz-2}
\eea where  the functions $a_{\pm}^{(s,s')}(u|\e_1,\e_2,\e_3)$ and
$d_{\pm}^{(s,s')}(u|\e_1,\e_2,\e_3)$ are given by (\ref{a-function-1}) and (\ref{d-function-1}) and
the $Q$-functions $Q^{(\pm)}(u)$ are  respectively, \bea
&&Q^{(\pm)}(u)=\prod_{j=1}^{M^{\pm}}\sinh(u-\l^{\pm}_j)\,\sinh(u+\l^{\pm}_j+\eta),
 \no \\[6pt] &&  M^{\pm}=\frac{1}{2}(2(s+s')N-1\mp k).\label{Q-2} \eea The
parameters $\{\l_j^{\pm}\}$ have to satisfy the conventional BAEs \bea
 \frac{a^{(s.s')}_{\pm}(\l^{\mp}_j|\e_1,\e_2,\e_3)}{d^{(s.s')}_{\pm}(\l^{\mp}_j|\e_1,\e_2,\e_3)}=
 -\frac{Q^{(\pm)}(\l^\pm_j+\eta)}{Q^{(\pm)}(\l^\pm_j-\eta)},\quad j=1,\ldots, M^{\pm}.\label{BAE-s-2}
\eea

\section{Conclusions}\setcounter{equation}{0}
The XXZ alternating spin chain with the generic non-diagonal
boundary terms specified by the most general non-diagonal
$K$-matrices given by (\ref{fusedKmatrix0103})-(\ref{K-0102}) is
studied by the off-diagonal Bethe anstz method.  Based on the
intrinsic properties (\ref{Int-R-ll})-(\ref{Fusion-Con-3}) of the
fused $R$-matrices  and $K$-matrices, we obtain the closed operator
identities (\ref{Operator-identity-1}) and
(\ref{Operator-identity-2}) of the fundamental transfer matrix
$t^{(\frac{1}{2})}(u)$. These identities, together with other
properties (\ref{Operator-perio})-(\ref{Operator-Asy}), allow us to
construct the off-diagonal (or inhomogeneous) $T-Q$ equation
(\ref{T-Q-ansatz-1})  and the associated BAEs (\ref{BAE-s-1})
accounting for the eigenvalues of the transfer matrix.

We remark that if the anisotropic parameter $\eta$ takes the
following discrete values \bea
\eta=-\frac{\a_-+\b_-+\a_++\b_+\pm(\theta_--\theta_+)+2i\pi
m}{2(s+s')N+1},\quad m\in Z, \eea our inhomogeneous $T$-$Q$ relation
(\ref{T-Q-ansatz-1}) can be reduced to the conventional one \bea
 \L^{(\frac{1}{2})}(u)&=&a^{(s,s')}(u)\frac{Q(u-\eta)}{Q(u)}
                           +d^{(s,s')}(u)\frac{Q(u+\eta)}{Q(u)}, \label{T-Q-ansatz-3}
\eea where the $Q$-function is given by (\ref{Q-function}). The
associated BAEs thus read \bea
\frac{a^{(s.s')}(\l_j)}{d^{(s.s')}(\l_j)}=
 -\frac{Q(\l_j+\eta)}{Q(\l_j-\eta)},\quad j=1,\ldots, 2(s+s')N.\label{BAE-s-3}
\eea When taking the thermodynamical limit $N\rightarrow \infty$
\footnote{At the same time, we also take the limit $m\rightarrow
\infty$ such that $\eta$ be finite.}, $\eta$ becomes dense on the
imaginary line. This allows one to use the method developed in
\cite{lcysw14} to study  the thermodynamic properties (up to the
order of $O(N^{-2})$) of the model for generic values of $\eta$ via
the conventional thermodynamic Bethe Ansatz methods \cite{Kor93}.

\section*{Acknowledgments}

We would like  thank R.\,I. Nepomechie  for his valuable
discussions. The financial support from  the National Natural
Science Foundation of China (Grant Nos. 11174335, 11375141,
11374334, 11434013, 11425522), the National Program for Basic Research of MOST
(973 project under grant No. 2011CB921700) and BCMIIS are gratefully
acknowledged. Two of the authors (W.-L. Yang and K. Shi) would like
to thank IoP, CAS for the hospitality. W. -L. Yang would also like
to thank KITPC for the hospitality where some part of the work was
done during his visiting.


\section*{Appendix A: $U_q(sl_2)$ algebra and spin-$l$ representation }
\setcounter{equation}{0}
\renewcommand{\theequation}{A.\arabic{equation}}
The underlying algebra of the XXZ spin chain is the quantum  algebra
$U_q(sl_2)$ generated by $\{S^{\pm},\,S^3\}$ with the relations
\cite{Cha94,fusion,fusion-1} \bea
&&[S^+,\,S^-]=\frac{\sinh(2\eta\,S^3)}{\sinh\eta},\quad
[S^3,\,S^{\pm}]=\pm S^{\pm}.\label{Commututing-relations} \eea The
Casimir operator $C_2$ of $\U$, which commutes with all the
generators,   is \bea
 C_2&=&\cosh(\eta+2\eta S^3)+2\sinh^2\eta\,\,S^-\,S^+  \no \\[6pt]
&=&\cosh(\eta-2\eta S^3)+2\sinh^2\eta\,\,S^+\,S^-.\label{Casi-1}
\eea All the generators of $\U$ can be realized by
$(2l+1)\times(2l+1)$ matrices on $(2l+1)$-dimensional spin-$l$ space
as follows \bea
&&S^3|m\rangle=m\,|m\rangle,\label{Rep-1}\\[6pt]
&&S^+|m\rangle=\sqrt{[l+1+m]_q[l-m]_q}\,|m+1\rangle,\label{Rep-2}\\[6pt]
&&S^-|m\rangle=\sqrt{[l+m]_q[l+1-m]_q}\,|m-1\rangle,\quad
m=l,l-1,\ldots,-l,\label{Rep-3} \eea where
$\{|m\rangle|m=l,l-1,\ldots,-l\}$ is an orthonormal basis of the
spin-$l$ space and we have used the notation \bea
[x]_q=\frac{q^x-q^{-x}}{q-q^{-1}},\quad q=e^{\eta}.\no \eea he
Casimir operator (\ref{Casi-1}) acting on the spin-$l$ space is
proportional to the identity operator, i.e., \bea C_2\,
|m\rangle=\cosh(\eta+2l\eta)\,|m\rangle,\quad
m=l,l-1,\ldots,-l.\label{Casi-2} \eea For the simplest case, i.e.,
$l=\frac{1}{2}$, the corresponding generators can be realized by
$2\times 2$ matrices \bea S^{\pm}=\s^{\pm},\quad
S^3=\frac{1}{2}\s^z,\quad C_2=\cosh2\eta\times{\rm id}.\no \eea
Moreover $U_q(sl_2)$ is a Hopf algebra with the following coproduct
$\Delta$ \cite{Cha94} \bea
&&\Delta(S^{3})=S^3\otimes {\rm id}+{\rm id}\otimes S^3,\label{Coproduct-1}\\[6pt]
&&\Delta(S^{\pm})=S^{\pm}\otimes q^{-S^3}+q^{S^3}\otimes
S^{\pm}\label{Coproduct-2}, \eea which is an algebraic homomorphism
and allows one to construct the representation on the tensor product
of two representation spaces.


\section*{Appendix B: Proofs of (\ref{Int-R-ll}) and (\ref{Fusion-Con-3})  }
\setcounter{equation}{0}
\renewcommand{\theequation}{B.\arabic{equation}}
We restrict to the spin-$l$ space ${\rm \bf V}$ (i.e., a
$(2l+1)$-dimensional vector space endowing a representation of
$U_q(sl_2)$ with the realization given by
(\ref{Rep-1})-(\ref{Rep-3})). The $R$-matrix $\bar{R}^{(l,l)}(u)\in
{\rm End}({\rm\bf V}\otimes {\rm\bf V})$ of $U_q(sl_2)$ with the
coproduct given by (\ref{Coproduct-1})- (\ref{Coproduct-2}) was
given in \cite{Jim85} (for the rational case \cite{Kul81}), \bea
\hspace{-1cm} \check{\bar{R}}^{(l,l)}(u)\equiv
\mathbf{P}\bar{R}^{(l,l)}(u)=\prod_{k=1}^{2l}\sinh(-u+k\eta)\,\sum_{m=0}^{2l}\prod_{k=1}^m\frac{\sinh(u+k\eta)}{\sinh(-u+k\eta)}\,
\bar{\mathbf{P}}^{(m)},\label{R-ll} \eea where $\mathbf{P}$ is the
permutation operator between the tensor product space of the
spin-$l$ vector spaces, $\bar{\mathbf{P}}^{(m)}$ is a projector
acting on the tensor product space of two spin-$l$ spaces by the
coproduct (\ref{Coproduct-1})-(\ref{Coproduct-2}) and projects the
tensor space into the irreducible subspace of spin-$m$ (i.e.,
$(2m+1)$-dimensional subspace). In particular, the $R$-matrix
$\bar{R}^{(l,l)}(u)$ also satisfies  the following important
properties \bea &&\hspace{-1.5cm}\mbox{ Initial
condition}:\,\bar{R}^{(l,l)}_{12}(0)= \prod_{k=1}^{2l}\sinh(k\eta)\,\mathbf{P}_{12},\label{Int-R-ll-1}\\[6pt]
&&\hspace{-1.5cm}\mbox{ Fusion
condition}:\,\bar{R}^{(l,l)}_{12}(-\eta)=\prod_{k=2}^{2l+1}\sinh(k\eta)\,\mathbf{P}_{12}\,\bar{\mathbf{P}}^{(0)}_{12}(l).
\label{Fusion-Con-4} \eea The projector $\bar{\mathbf{P}}^{(0)}(l)$
projects the tensor space of two spin-$l$ spaces to the singlet
state, namely, \bea
\bar{{\mathbf{P}}}^{(0)}(l)=\frac{|\bar{\Phi}^{(l)}_0\rangle\langle\bar{\Phi}^{(l)}_0|}{\langle\bar{\Phi}^{(l)}_0|\bar{\Phi}^{(l)}_0\rangle},\quad
|\bar{\Phi}^{(l)}_0\rangle=\sum_{k=0}^{2l}(-q)^k|l-k\rangle\otimes|-l+k\rangle.\label{Singlet-1}
\eea The singlet state $|\bar{\Phi}^{(l)}_0\rangle$ of $U_q(sl_2)$ is the
unique vector (up to a rescaling) which enjoys the following
properties \bea
\Delta(S^3)|\bar{\Phi}^{(l)}_0\rangle=\Delta(S^{\pm})|\bar{\Phi}^{(l)}_0\rangle=0.
\eea It was shown in \cite{Byt03} that the $R$-matrices
$R^{(l,l)}(u)$ and $\bar{R}^{(l,l)}(u)$ have the following relation
\bea
 R^{(l,l)}(u)=\lt({\rm id}\otimes e^{u S^3}\rt)\bar{R}^{(l,l)}(u) \lt({\rm id}\otimes e^{-u S^3}\rt).\label{R-matrix-relation}
\eea The above relation and the properties
(\ref{Int-R-ll-1})-(\ref{Singlet-1}) imply that  the fused
spin-$(l,l)$ $R$-matrix $R^{(l,l)}(u)$ given by (\ref{fused-Rjl})
with $j=l$ indeed  satisfies  the  very relations (\ref{Int-R-ll})
and (\ref{Fusion-Con-3}).

\end{document}